\documentstyle[prl,tighten,aps,epsf,preprint]{revtex}
\begin{document}   
\title{Rigidity of Orientationally Ordered Domains of
Short Chain Molecules}
\author{Hiroaki Nakamura, Susumu Fujiwara and Tetsuya Sato}
\address{Theory and Computer Simulation Center,
National Institute for Fusion Science,
322-6, Oroshi-cho, Toki-shi, Gifu,  509-5292, Japan \\
(\today)}
\maketitle
\begin{abstract} 
By molecular dynamics simulation, discovered is a strange rigid-like nature for a hexagonally packed domain of short chain molecules. In spite of the non-bonded short-range interaction potential (Lennard-Jones potential)  among chain molecules, the packed domain gives rise to a resultant global moment of inertia. Accordingly, as two domains encounter obliquely, they rotate so as to be parallel to each other keeping their overall structures as if they were rigid bodies.
\end{abstract}

Recent molecular dynamics (MD) simulations
by Fujiwara and Sato\cite{98f,99f,99f2} have
demonstrated that randomly agitating short chain molecules immersed 
in a hot heat bath are packed into an orientationally-ordered hexagonal structure 
by sudden cooling to a low temperature.
In this evolution process, ``domains" of orientationally ordered chain molecules first 
grow randomly and locally at different places and then, 
as neighboring domain boundaries encounter, they coalesce rapidly 
with each other
and develop into one  large hexagonal structure  in a stepwise fashion.
 
Very recently, an experimental study on the structure formation of {\it s}-polypropylenes
by Heck {\it et al.} \cite{99h} has also disclosed that crystal blocks are first emerging
in planner assemblies and then merge to each other to form a lamella structure.

These simulation and experiment results are interesting enough to challenge us to reveal how locally formed domains or blocks coalesce.
The previous Fujiwara-Sato (FS) simulation indicates that the coalescence of 
two domains take place almost instantaneously during the relaxation process
from a random configuration to an orientationally-ordered hexagonal structure.

In the model of the FS simulation,
the temperature of the heat bath in which chain molecules were immersed was fixed 
to a finite value (400 K), so that each chain molecule suffered from thermal fluctuation. 
This fluctuation makes it difficult to investigate motion of the chain molecules 
under the Lennard-Jones (LJ) potential interaction.
We, therefore, keep the temperature of the system $T=0$ K 
to remove the effect of thermal fluctuation 
and place two hexagonally packed domains in contact side by side 
with a certain tilt angle. 
With this initial condition we perform MD simulations 
and examine the dynamical evolution of the two domains.
We adopt the united atom approximation.
The united atoms interact via the bonded potentials 
(bond-stretching, bond-bending, and torsional potentials) and
the non-bonded potential (LJ potential).
The atomic force field used here is the DREIDING potential\cite{90m}.
The numerical integration of the equations of motion is
performed using the velocity Verlet algorithm \cite{82s}.
The integration time step is 1.0 fs. 
The cutoff distance for LJ potential is 10.5 \AA. 
The units of length and energy are $\sigma = 0.36239 $ nm and 
$\epsilon = 0.1984 $ kcal/mol, respectively.
A chain molecule consists of 20 united atoms; the mass of the united atoms is $m=14 $ g/mol.

The MD simulation is carried out by the following procedure.
We allocate chain molecules in all-{\it trans} states 
and pack such 61 chains hexagonally.
The distance between the nearest chains is 0.75 nm;
this value  was chosen from the previous work\cite{98f}.
We prepare a pair of hexagonal domains,
each consisting of 61 chain molecules, and put side by side in such a way
that the two domains are in contact at one side of a hexagonal cylinder.
Then, in  order to find a minimum energy configuration, 
we run MD simulation for some period with the heat bath temperature of 
$T = 5 $ K,
until the physical quantities, such as  LJ potential energy and bonded-potential 
energy, show saturation.
Finally, we reduce the temperature of the system gradually from $T= 5$ K to 
0 K and run simulation at $T=0$ K to obtain a minimum energy state.
The Nos\'e-Hoover method\cite{84n,85h}
is used in the above simulation to
keep the temperature of the system constant.

With this preparation of initial setup of the simulation environment,
we stand on the starting line of our main simulation run.
We rotate one domain by $\pi/6$ against the other domain and 
start simulation. 
We note here that the Nos\'e-Hoover method is not used but only equations of motion are solved in this simulation run.

\begin{figure}
\epsfxsize=\textwidth \epsfbox{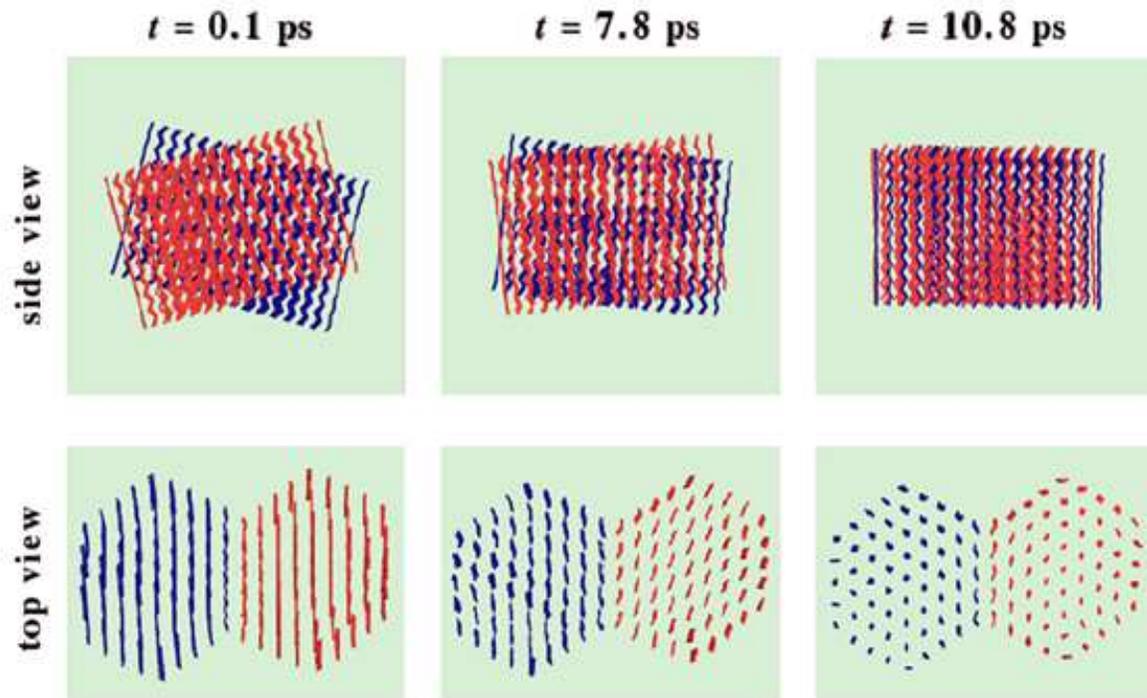}%
  \caption{Structural evolution of two contact domains with a tilt angle $\pi/6$. 
  Two domains are, respectively, colored by red and blue. }
  \label{dom.fig}
\end{figure}

The overall evolution of the two domains are shown in Fig.~\ref{dom.fig},
where the top and bottom panels are respectively the top and side views.
From Fig.~\ref{dom.fig} it is seen that two domains start moving 
and their orientations become almost parallel at 10.8 ps, as if they were rigid bodies.
From this fact, we expect that the LJ potential must have produced an overall torque 
to make the domains parallel to each other.

The observation that two domains behave as if they were rigid bodies 
is a striking fact. 
This is because the interaction forces connecting
different chains are the short-range forces of the non-bonded LJ potential 
under which only the nearest chains interact predominantly.
However, it appears that all chains in the same domain react collectively
and each domain behaves as if it were a rigid body.

To reveal this striking and remarkable behavior, namely, rigidity of packing, 
we investigate the time evolutions of the orientations of chain molecules and domains.
For this purpose, we define each direction vector of chain molecule in the right domain 
as $ {\mbox{\boldmath $u$}}^{\rm R}_i (t) (i=1, \cdots , 61 ),$ 
which is a unit vector of the principal axis 
with the smallest moment of inertia of the $i$th chain at time $t$.
The direction vector of the right domain, 
$  {\mbox{\boldmath $U$}}^{\rm R} (t)$  is 
also defined  by the average of all chain-molecules' direction-vectors belonging to the right domain,
as follows:
\begin{equation}
  {\mbox{\boldmath $U$}}^{\rm R} (t)
  \equiv \frac{\sum_{i=1}^{61 } 
       { \mbox{\boldmath $u$} }^{ \rm R }_i (t)}{
        | \sum_{i=1}^{ 61  } 
       { \mbox{\boldmath $u$} }^{ \rm R }_i (t)| }. \label{eq.1}
 \end{equation}
For the left domain, 
we define the directions of each chain and the left domain
by changing the suffix ``R" to ``L", which denotes the left domain,
in the above definitions.

\begin{figure}
\epsfxsize=8cm \epsfbox{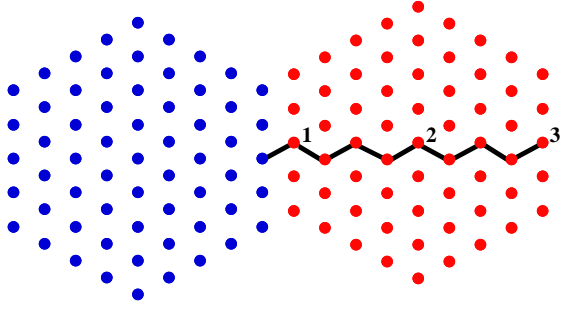}%
  \caption{Top view of schematic configuration of (61+61) chains 
 before tilting.
 A solid zigzag line denotes the shortest path of the information transfer
 from the contact point to the edge.
 } \label{chain}
\end{figure}

\begin{figure}
\epsfxsize=8cm \epsfbox{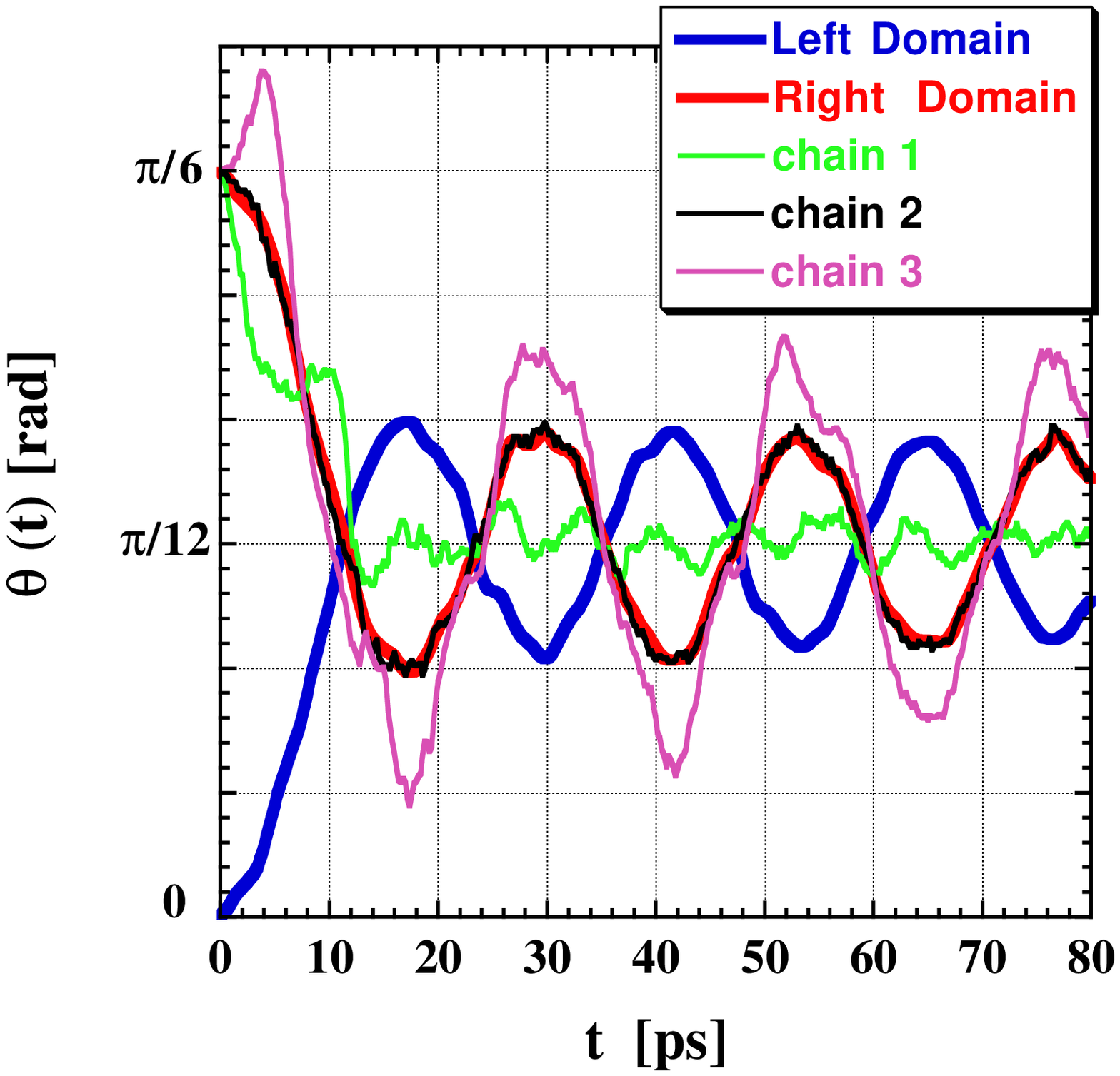}
\caption{
Time evolution of the angles of
$ {\mbox{\protect \boldmath $U$}}^{\rm L} $ (blue),
$ {\mbox{\protect \boldmath $U$}}^{\rm R} $ (red), 
$ {\mbox{\protect \boldmath $u$}}^{\rm R}_1 $ (green),
$ {\mbox{\protect \boldmath $u$}}^{\rm R}_2 $ (black), and
$ {\mbox{\protect \boldmath $u$}}^{\rm R}_3 $ (purple)
 (Chain 1, 2 and 3 are indicated in Fig.~\protect\ref{chain}).
} \label{theta.fig}
\end{figure}

We pick up three chain molecules in the right domain, which are indicated 
by 1, 2 and 3 in Fig.~\ref{chain}.
Calculating each angle $\theta$ between each chain's direction vector and the basis vector 
$ \mbox{ \boldmath $e$ } ( \equiv {\mbox{\boldmath $U$}}_{\rm L} (0)) $, 
the time dependence of the angle is plotted in Fig.~\ref{theta.fig}.
Blue and red solid curves denote the time evolutions of the left and right domain angles, respectively.
It is found that the domain relaxation time is of the order of 10 ps.
Figure \ref{theta.fig}
shows that there appears some  time delay $\delta t$ 
between the angle of the chain on the contact surface 
and that of the chain on the edge;
the value of $\delta t$ is estimated to be 4 ps from this figure.
This fact gives us an evidence that the domain motion is not ``rigid" in a strict sense 
but something like an ``elastic" body.
From our observations we come up with the following scenario: 
At first, a torque appears on the contact surface between the tilted domains 
and then it is transferred to the chain molecules one by one. 
It takes about 4 ps for information to reach 
from the contact surface to the edge of the domain.   
After this initial information exchange of about 4 ps,
the phase delay through the domain is mixed up
and all chain molecules oscillate in phase.
On observing this chain molecules' motion in the long time scale 
of the domain's motion, we find that 
the time delay of chain molecules is negligible.
Therefore, the motion of domain may be regarded as a ``quasi"-rigid body.

The binding interaction forces come from the LJ potential and the  three bonded
potentials (bond-stretching,  bond-bending, and torsional potentials).
Since the potential function of each force is known,
we can evaluate the oscillation period in each potential well around the minimum
energy state.
The period is given in Table. \ref{pot.tab}.
From this table we can say that the binding forces due to the bonded potentials
are harder than that of the LJ potential. 
This indicates that it would be hard to excite each bonded atom 
in a chain molecule and to destroy the bonded chain structure. 
In other words, we may well regard each chain as a rigid chain.
From this consideration, we can pay attention only 
to the rigidity of the LJ potentials acting upon the neighboring chains.

\begin{figure}
\epsfysize=7cm \epsfbox{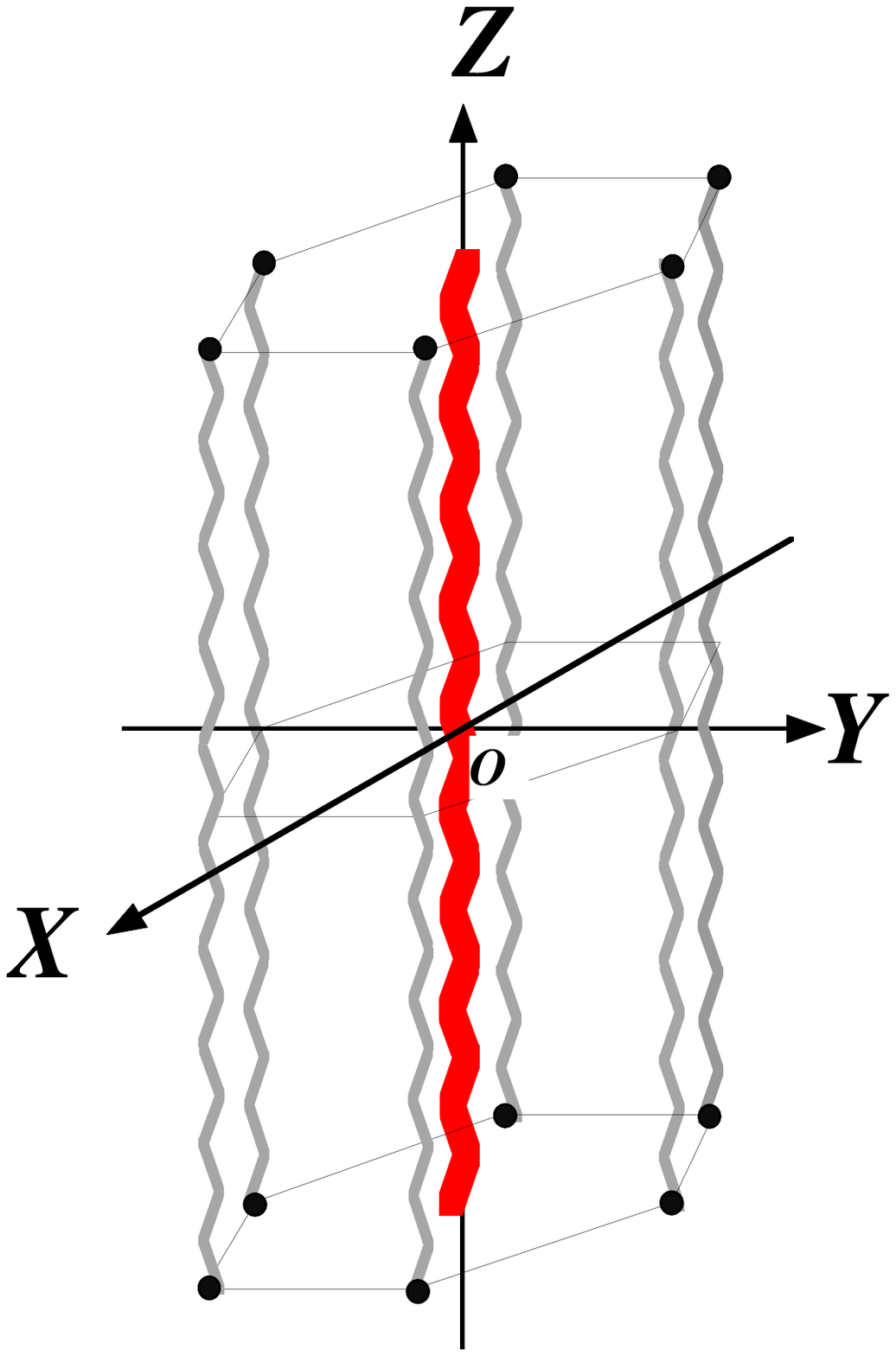}
\caption{ 
Schematic illustration of the unit hexagonal cell (7 rigid-chain model).
} \label{hex}
\end{figure}

\begin{figure}
\epsfxsize=8cm \epsfbox{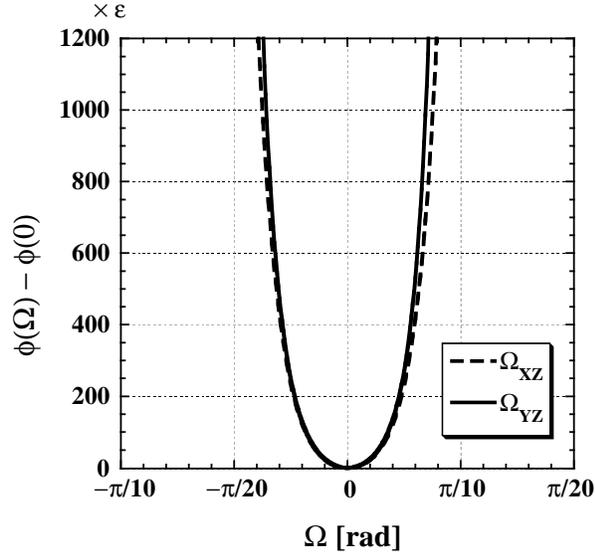}
\caption{The Lennard-Jones potential of the unit hexagonal cell
as a function of the  rotation angle $\Omega_{i} \ \ ( i = XZ, YZ )$.
Rotation is made for the central rod (red) in Fig.~\protect\ref{hex}
in the $XZ$ or $YZ$ plane around the origin.}   \label{cell.fig}
\end{figure}

In order to examine the rigidity of the LJ potentials in hexagonally packed domains,
we, first of all, pick up one unit hexagonal cell that consists of 7 chains, 
one chain being surrounded by hexagonally distributed 6 chains (the 7 rigid-chain model)
as shown in Fig.~\ref{hex}.
We can derive the LJ potential $\phi$ as a function 
of the tilt angle of the central chain $\Omega$;
the central chain consists of the 20 united atoms, the position of which 
are denoted by $ \mbox{\boldmath $r$}_{\rm s} (s=1,\cdots,20),$ and 
the center of mass of the central chain is defined 
by  $ \mbox{\boldmath $r$}_{\rm c}.$
The LJ potential $\phi (\Omega)$ is plotted in Fig.~\ref{cell.fig}
when the tilt is made in the $XZ$ plane and in the $YZ$ plane.
The valley of the potential $\phi$ is deeper 
and steeper than the LJ potential of one united atom. 
Therefore, when the central chain is deviated 
from the balanced position, a restoring force, 
which puts the molecule back to the balance point, 
is intensified 
compared with the case of randomly distributed chains.

To estimate quantitatively the hardness of the packed lattice, 
we calculate a restoring oscillation period $\tau$ of oscillation 
in the potential valley, $\phi(\Omega).$
To do so, we need physical values of the packed lattice.
In the vicinity of the minimum energy point, we may well 
approximate the LJ potential by the parabolic function as follows:
\begin{equation}
\phi(\Omega_i) -\phi (0)
\sim \frac{1}{2} k_{i} \Omega_{i}^2, 
\ \ (i=\mbox{$XZ, YZ$}) .
\label{eq.6} 
\end{equation}
The values of $k$'s are given in Table \ref{cell.tab}. 
The restoring period of the motion of a chain around the minimum energy state
in the parabolic potential $\phi$ is given by 
\begin{equation}
\tau_{i} = 2 \pi
        \sqrt{ \frac{  I }{ k_{i} }  },
\ \ (i=\mbox{$XZ, YZ$}) , 
 \label{eq.11} 
\end{equation}
where $I$ is the moment of inertia of the central chain molecule of
the hexagonal configuration around the 
center of mass   $ \mbox{\boldmath $r$}_{\rm c}$, i.e., 
$I = \sum_{s=1}^{20} m
     \left( \mbox{\boldmath $r$}_{s} -
               \mbox{\boldmath $r$}_{\rm c}
     \right)^2  \sim 79.1 m \sigma^2 .$
By plugging the $k$'s values given in Table \ref{cell.tab} 
and the obtained $I$ value into Eq. (\ref{eq.11}), 
the restoring period $\tau$ of the central chain is calculated to be about 0.36 ps
as given in Table \ref{cell.tab}.
This value is of the same order as the torsional bonded potential.
This fact indicates that the non-bonded  LJ potential among chain molecules 
is strengthened by hexagonal packing  to the same order as the torsional bonded potential.
Consequently, the non-bonded LJ potential acts to bind strongly neighboring chains
as if it were  the bonded potential.

By using the obtained $\tau$,
for the unit hexagonal cell, we shall next go on to estimate the resultant 
relaxation time of the whole domain.
Since the restoring information must pass over $9$ unit neighboring hexagonal cells
until the information reaches the edge of the domain from the contact surface,
the time delay is estimated to $9\times \tau \sim 3.2 $ ps.
This obtained value is of the same order as the simulation result, of 4 ps.
The difference between the values must be caused by 
the omission of the bonded potentials in the 7 rigid-chain model.

In conclusion, it is demonstrated that hexagonally packed chain molecules 
at low temperatures behave as if they were quasi-rigid bodies,
despite the fact that the packing is done preliminarily by the weak, non-bonded
short-range LJ potential.

This work was carried out by the Advanced Computing System
for Complexity Simulation (NEC SX-4/64M2) at National Institute
for Fusion Science.

\begin{table}
\caption{Oscillation periods for the interaction potentials.}
\label{pot.tab}
\begin{tabular}{c|c||c} 
 \multicolumn{2}{c||}{   } & oscillation period [ps] \\ \hline
                   & bond-stretching  & 0.03 \\ 
   bonded potentials                & bond-bending  & 0.12\\ 
                   & torsional  & 0.37 \\ \hline
  \multicolumn{2}{c||}{non-bonded potential (LJ potential)}  & 0.82 \\ 
\end{tabular}
\end{table}

\begin{table}
\caption{Coefficients $k$ of the Lennard-Jones potential 
approximated by the  parabolic function
and oscillation periods of the unit hexagonal cell.
 }
\label{cell.tab}

\begin{tabular}{ccc} 
& $k$ [kcal/mol$\cdot{\rm rad}^2$ ] & oscillation period [ps] \\ 
\tableline
                  $XZ$-plane & $8970 \pm 40 $  
                  & $0.3680 \pm (8\times 10^{-4})$ \\ 
       $YZ$-plane & $9290 \pm 40  $    
        &  $0.3617 \pm (7\times 10^{-4})$ \\ 
\end{tabular}
\end{table}

\clearpage

\newlength{\minitwocolumn}
\setlength{\minitwocolumn}{0.45\textwidth}
\noindent

\clearpage

\end{document}